\documentclass[onecolumn,showpacs]{revtex4}

\usepackage{graphicx}% Include figure files
\usepackage{dcolumn}% Align table columns on decimal point
\usepackage{bm}% bold math

\input epsf

\begin{document}

\title{\Large Gravitational collapse due to dark matter and dark energy in the brane world scenario}

\author{\bf Soma Nath$^1$,~Subenoy
Chakraborty$^1$\footnote{subenoyc@yahoo.co.in} and Ujjal
Debnath$^2$\footnote{ujjaldebnath@yahoo.com}}

\affiliation{$^1$Department of Mathematics, Jadavpur University,
Calcutta-32, India.\\ $^2$Department of Mathematics, Bengal
Engineering and Science University, Shibpur, Howrah-711 103,
India.\\}

\date{\today}

\begin{abstract}
Gravitational collapse of FRW brane world embedded in a conformaly
flat bulk is considered for matter cloud consists of dark matter
and dark energy with equation of state $p=\epsilon \rho$
$(\epsilon<-\frac{1}{3})$. The effect of dark matter and dark
energy is being considered first separately and then a combination
of them both with and without interaction. In some cases the
collapse leads to black hole in some other cases naked
singularity appears.

\end{abstract}

\pacs{04.20~-q,~~04.40~ Dg,~~97.10.~CV.}

\maketitle

\section{\normalsize\bf{Introduction}}

        In the last decade, two remarkable advancements has been done - one
of them is the theoretical idea of the brane world scenario
$[1,2]$ and the other one is the observational evidence 
that our universe is at present accelerating [3-5].\\

         Randall and Sundrum $[1]$ first proposed the bulk-brane model to
  explain the higher dimensional theory. According to them, we
  live in a four dimensional world (called $3$-brane, a domain wall) which
  is embedded in a $5$-dimensional space-time (called bulk). The
  extra fifth dimension may be finite or infinite in size. All
  matter fields are confined in the brane while gravity can only
  propagate in the bulk.\\

           The observational fact of accelerating universe was
  first noticed from high red shift supernova Ia [3]  and was
  confirmed from the cosmic microwave background radiation $[4]$ and
  large scale structure [5]. In Einstein  gravity present accelerating
  universe is possible with matter field (dark energy) having large
  -ve pressure (violating energy conditions). From observational
  point of view, our flat universe contains approximately 72$\%$
   of dark energy, 21$\%$ of dark matter, 4.5$\%$
  of visible matter (baryon matter) and 0.5$\%$
  radiation. As the nature of dark energy [6-11] is unknown so several
  models have been proposed for it such as Chaplygin gas,
  phantoms, quintessence, a tiny +ve cosmological constant, dark
  energy in brane worlds.\\

Recently, Cai and Wang $[12]$ have studied an important topic in
  gravitational physics namely black-holes and their formation in
  the universe considering gravitational collapse of dark matter
  in the background of dark energy. In the present work, formation
  of black-hole will be studied for brane world scenario with dark
  matter having dark energy in the background. The paper is
  organized as follows: in section II the basic equations for the
  gravitational
  collapse of a homogeneous and isotropic brane model having
  finite radius will be presented. Section III will deal with collapsing
  process for dark energy and dark matter separately to identify
  their individual role. In section IV, the collapse of dark
  matter in he presence of dark energy both without and with
  interaction between them (except the gravitational one) will be
  studied. The paper will end with  conclusions in section V.\\\\

\section{\normalsize\bf{General Formulation Of The Collapsing
process In Brane World }}

Suppose in the $5$D space-time (bulk) the brane-world is located
at $\chi(x^{\mu})=0$ with $x_{\mu},\mu=0,1,2,3,4$ are the $5$D
co-ordinates. The effective action in 5D [13] is
\begin{equation}
A=\int d^{5}x \sqrt{-^{(5)}g }~\left[\frac{1 }{2 \kappa ^{2}_{5}}
^{(5)}R-^{(5)}\Lambda \right]+ \int _{\chi=0} d^{4} x \sqrt{-
^{(4)}g} \left[\frac{1}{\kappa ^{2}_{5}} \kappa ^{\pm} -\lambda +L
^{matter}\right]
\end{equation}
Here $\kappa ^{2}_{5}=8 \pi ^{(5)} G$ is the $5$D coupling
constant, $\kappa ^{\pm}$ is the extrinsic curvature on either
side of the brane, $\lambda$ is the vacuum energy on the brane
(brane tension), $^{(5)}\Lambda $ is the -ve vacume energy, the
only source of the gravitational field in the bulk and
$L^{matter}$ is the Lagrangion  for the matter field on the brane.\\

The variation of the action gives the Einstein equation on the
bulk as
\begin{equation}
G_{\mu \nu}=\kappa ^{2}_{5}\left[-^{(5)}\Lambda g_{\mu \nu}+
\delta (\chi)(-\lambda g_{\mu \nu} +T ^{matter}_{\mu \nu})\right]
\end{equation}
Now, assuming $z_{2}$ -symmetry, the effective four dimensional
gravitational equations on the brane take the form [2,14]
\begin{equation}
G_{ij}=-^{(4)}\Lambda   g_{ij} +\kappa ^{2}_{4}   T_{ij} +\kappa
^{4}_{5}   \Pi _{ij} -E_{ij}
\end{equation}
where,
$$\Pi _{ij}=\frac{1}{12} T T_{ij}- \frac{1}{4}
T^{k}_{i}T_{jk}+\frac{1}{24} g_{ij}(3 T ^{lm}T _{lm} -T^{2})$$
 is the quadratic correction term,
 $E_{ij}$
 is the electric part of the $5$D Weyl tensor,
 $$^{(4)}\Lambda =\frac{\kappa ^{2}_{5} }{2}\left(^{(5)}\Lambda + \frac{\kappa ^{2}_{5} \lambda^{2}
 }{6}\right)$$
is the effective 4D cosmological constant and $$\kappa
^{2}_{4}=\kappa ^{4}_{5} \frac{\lambda}{6 } $$ is the
$4D$ gravitational coupling constant. \\

In the following assuming flat FRW space time
$$ds^{2}=dt^{2}+a^{2}(t)[dr^{2} +r^{2} (d\theta^{2} +Sin^{2}\theta
d\phi^{2} )]$$ as our brane world embedded in a conformally flat
bulk geometry with vanishing $^{(4)}\Lambda $, the effective
Einstein equations become
\begin{equation}
\frac{\dot{a^{2}}}{a^{2}}=\frac{\kappa}{3}[\rho_{T}
(1+\frac{\rho_{T}}{2 \lambda})]
\end{equation}
\begin{equation}
\frac{\ddot{a}}{a}=-\frac{\kappa}{6}[\rho_{T} (1+\frac{2\rho_
 {T}}{ \lambda}) +3 \varepsilon \rho (1+\frac{ \rho_
{T}}{\lambda})]
\end{equation}
with $\rho_{T}=\rho_{M}+\rho$. Here the energy momentum tensor
$T_{ij}$ is given by
\begin{equation}
T_{ij}=(\rho _{M}+\rho+ p) u_{i} u_{j} -p g_{ij}
\end{equation}
where $\rho_{M}$ stands for the energy density of the dark matter,
while $\rho$ and $p$ are respectively the energy density and
pressure of the dark energy having equation of state
$$p=\varepsilon \rho$$ and $u _{i}$ is their four-velocity having
expression $$u_{i}=\delta ^{t}_{i}$$ in the comoving co-ordinate
system. Using the energy-momentum conservation relation$$T _{ij;
k}~g^{jk}=0,$$ the interaction $Q(t)$ between dark matter and dark
energy can be expressed as
\begin{equation}
\dot{\rho}_{M}+3 \frac{\dot{a}}{a} \rho_{M}=Q
\end{equation}
\begin{equation}
\dot{\rho}+ 3 \frac{\dot{a}}{a}(1+\varepsilon) \rho=-Q
\end{equation}
Now, if we consider gravitational collapse of a spherical cloud
consists of above dark matter and dark energy distribution and is
bounded by the surface $\Sigma : r=r_{\Sigma}$ then the metric on
it can be written as
\begin{equation}
ds^{2}=dT^{2}-R^{2} (T) (d\theta^{2} +Sin^{2} \theta d\phi^{2}))
\end{equation}
Thus on $\Sigma : T=t$ and $R(T)=r_{\Sigma} a(T)$ where $
R(r,t)\equiv r a(t)$ is the geometrical radius of the two spheres
$t,r =$ constant. Also the total mass of the collapsing cloud is
given by
\begin{equation}
M(T) = m(r,t)\mid_{r=r_{\Sigma}}=\frac{1}{2} r^{3} a
\dot{a}^{2}\mid _{\Sigma}=\frac{1}{2}R(T)\dot{R}^{2}(T)
\end{equation}
The apparent horizon is defined as [15]
\begin{equation}
R,_{\alpha} R,_{\beta} g ^{\alpha \beta}=0,~~{\text i.e.,}~~
r^{2}\dot{a}^{2}=1
\end{equation}
So if $T=T_{AH}$ be the time when the whole cloud starts to be
trapped then
\begin{equation}
\dot{R}^{2}(T_{AH})\mid_{\Sigma}
=r_{\Sigma}^{2}\dot{a}^{2}(T_{AH})=1
\end{equation}
As it is usually assumed that the collapsing process starts from
regular initial data so initially at $t=t_{i} ~(<T _{AH})$, the
cloud is not trapped i.e.,
\begin{equation}
r_{\Sigma}^{2} \dot{a}^{2}(t_{i})< 1,~~~~
 (r_{\Sigma}\dot{a}(t_{i})>-1)
\end{equation}
 Thus if equation $(12)$ has any real solution for
$T_{AH}$ satisfying (13) then black hole will form, otherwise
the collapsing process leads to a naked singularity.\\

\section{\normalsize\bf{Role Of Dark Matter Or Dark Energy In Gravitational Collapse}}
This section will show the role of dark matter and dark energy
separately during the collapse.\\

{\bf CASE ~I : Collapse With Dark Matter: $\rho _{M}\neq 0, ~\rho
=p =0$}\\

 Here from conservation equations $(7)$ and $(8)$, on
integration one gets
\begin{equation}
\rho_{M}=\frac{\rho_{0}}{a^{3}}
\end{equation}

\begin{figure}
\includegraphics[height=2.7in]{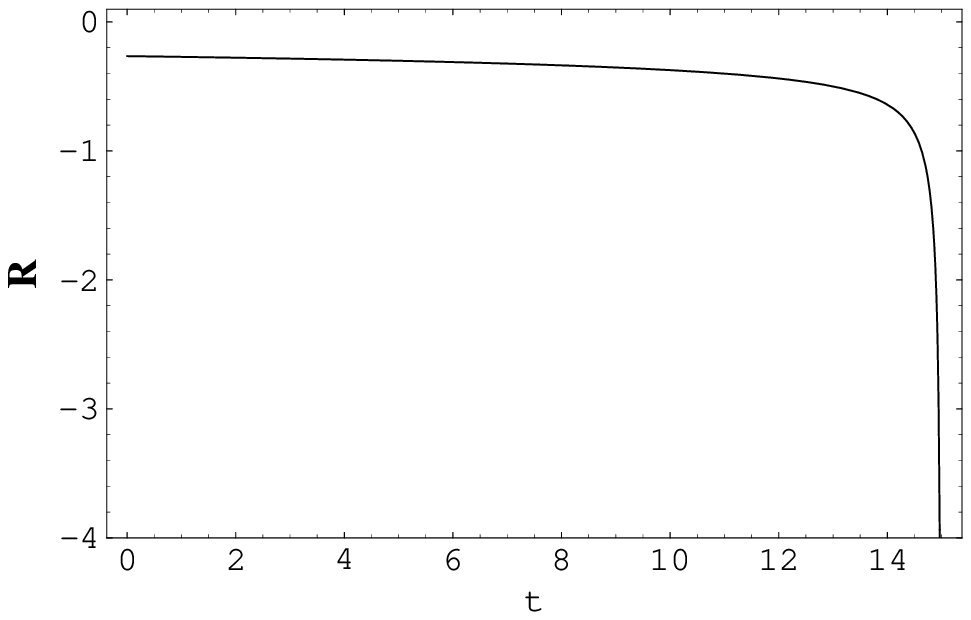}\\
\vspace{1mm} Fig.1\\

\vspace{5mm} Fig. 1 shows the behaviour of $\dot{R}$ of eq.(16) as
the collapsing process approaches to singularity. \hspace{2cm}
\vspace{4mm}
\end{figure}

Then using $\rho_{M}$ in equation $(4)$ and integrating one
obtains
\begin{equation}
a^{3} (t)=[a_{0} (t_{0}-t) +A_{0}]^{2}-A_{0}^{2}
\end{equation}
where $a_{0}=(\frac{3 \kappa
\rho_{0}}{4})^{\frac{1}{2}}$,~$A_{0}=(\frac{\rho_{0}}{2
\lambda})^{\frac{1}{2}}$ and $t_{0}$ is an integration constant.
The expression for relevant physical quantities are

\begin{equation}\left.
\begin{array}{llll}
\dot{R}(T)=-\frac{2}{3}R_{0}\frac{(a^{3} +
A_{0}^{2})^{\frac{1}{2}}}{a^{2}}
\\\\
 M(T)=\frac{2}{9}R_{0}^{3}\frac{(a^{3} +
A_{0}^{2}))}{a_{0} a^{3}}
\end{array}\right\}
\end{equation}
with $R_{0}=r_{\Sigma} a_{0}$.\\

 The above solution shows that as $t\rightarrow \infty,
a\rightarrow \infty, \rho_{M}\rightarrow 0,\dot{R}\rightarrow 0$
and $M(T)\rightarrow \frac{2 R_{0}^{3}}{9 a_{0}}$  . So the
collapsing process (see fig.1) with dark matter starts at
$t=t_{i}$ (say) when the condition $(13)$ is satisfied and the
time of formation of apparent horizon $(T_{AH})$ is given by the
real root of the equation
\begin{equation}
[a_{0}^{2}(T_{0}-T)^{2}+2a_{0}A_{0}(T_{0}-T)]^{4}=\left(\frac{4
R_{0}^{2}}{9}\right)^{3}[a_{0}(T_{0}-T)+A_{0}]^{6}
\end{equation}
Finally, the singularity will form at $T=T_{0}$. Note that during
the collapse  the dark matter energy density and the total mass
of the cloud gradually increase and go to infinity at
the singularity.\\

{\bf CASE II : Collapse With Dark Energy:
$\rho_{M}=0,~p=\varepsilon \rho$}\\

In this case the collapsing cloud is in the form of perfect fluid
obeying barotropic equation of state : $p=\varepsilon \rho$,
$\varepsilon$ is a constant. It is to be noted that when
$\varepsilon < -\frac{1}{3}$, the strong energy condition is
violated and the fluid is termed as dark energy. The solution for
different values of $\varepsilon$ are given by\\\

\begin{figure}
\includegraphics[height=1.7in]{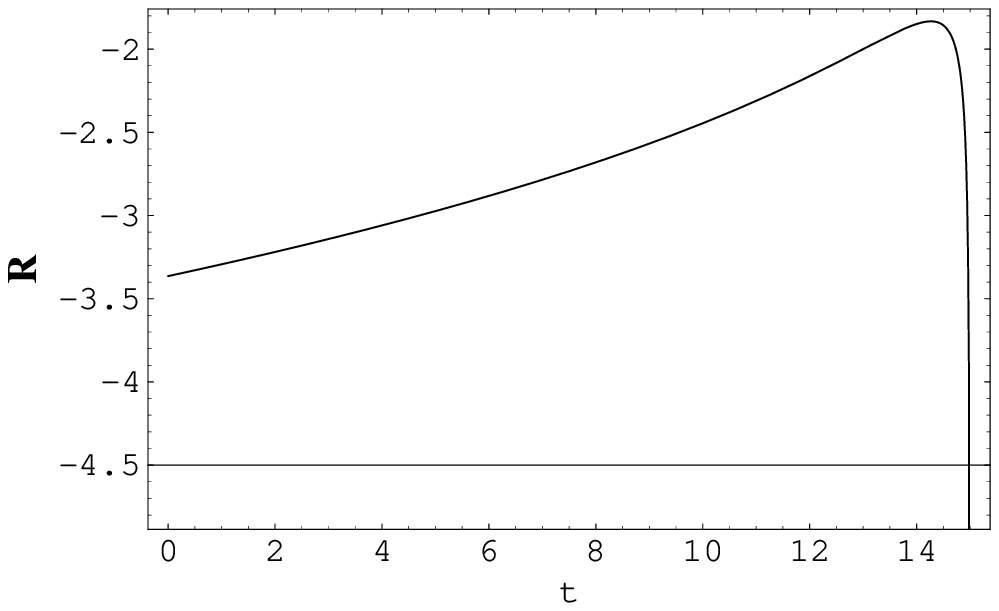}~~~~
\includegraphics[height=1.7in]{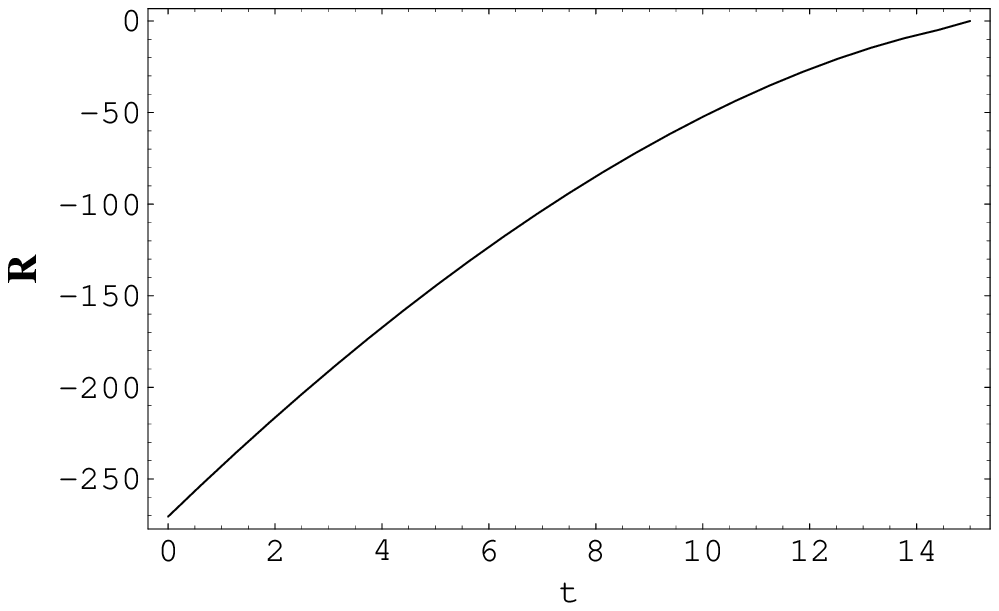}\\
\vspace{1mm} Fig.2~~~~~~~~~~~~~~~~~~~~~~~~~~~~~~~~~~~~~~~~~~~~~~~~~~~~~~~~~~~~~~~~~~~ Fig.3\\

\vspace{5mm} Fig. 2 and 3 show the variations of $\dot{R}$ with
time in eq. (18) for $\varepsilon=-1/2$ and $\varepsilon=-3/4$
respectively. \hspace{2cm} \vspace{4mm}
\end{figure}

\begin{equation}\left.
\begin{array}{llll}
a^{3(1+\varepsilon)}=[a_{1} (t_{0}-t)+A_{1}]^{2}-A_{1}^{2}
\\\\
\rho=\frac{\rho_{1}}{a^{3(1+\varepsilon)}}
\\\\
\dot{R}(T)=-\frac{2
R_{1}[a^{3(1+\varepsilon)}+A_{1}^{2}]^{\frac{1}{2}}}{3
(1+\varepsilon)a^{3(1+\varepsilon)-1}}
\\\\
M(T)=\frac{2
R_{1}^{3}[a^{3(1+\varepsilon)}+A_{1}^{2}]}{9(1+\varepsilon)^{2}a_{1}a^{6(^{1+\varepsilon})-3}}
\end{array}\right\}~~~~~~~~\varepsilon>-1
\end{equation}\\\

\begin{equation}\left.
\begin{array}{llll}
a=a_{2} e^{\mu (t_{0}-t)}
\\\\
\rho=\rho_{1}
\\\\
\dot{R}(T)=-R_{2}\mu e^{\mu (t_{0}-t)}
\\\\
M(T)=\frac{R_{2}^{3}\mu^{2}}{2} e^{3\mu (t_{0}-t)}
\end{array}\right\}~~~~~~~~\varepsilon =-1
\end{equation}\\\
and
\begin{equation}\left.
\begin{array}{llll}
a=[{a_{3}(t-t_{0})}^{2} -A_{1}]^{-\frac{1}{3(|\varepsilon| -1)}}
\\\\
\rho=\frac{\rho_{1}}{{a_{3}(t-t_{0})}^{2} -A_{1}}
\\\\
\dot{R}(T)=-\frac{2}{3}R_{0}^{'}(t-t_{0})[{a_{3}(t-t_{0})}^{2}
-A_{1}]^{\frac{2-3|\varepsilon|}{3(|\varepsilon|-1)}}
\\\\
M(T)=\frac{2}{9}\frac{R_{0}^{' 3}
(T-T_{0})^{2}}{a_{0}^{'}[a_{3}(T-T_{0})^{2}-A_{1}]^{\frac{2
|\varepsilon|-1}{|\varepsilon|-1}}}
\end{array}\right\}~~~~~~~~\varepsilon< -1
\end{equation}

where the expressions for different constants are
$$A_{1}=\sqrt{\frac{\rho_{1}}{2 \lambda}},~a_{1}=\frac{\sqrt{3\kappa
\rho_{1}}}{2}(1+\varepsilon),~R_{1}=r_{\Sigma}a_{1},~\mu
=[\frac{\kappa
\rho_{1}}{3}(1+A_{1}^{2})]^{\frac{1}{2}},~R_{2}=r_{\Sigma}a_{2},$$$$a_{3}=\frac{\sqrt{3\kappa
\rho_{1}}}{2}(|\varepsilon|-1),~a_{0}^{'}=\frac{\sqrt{3\kappa
\rho_{1}}}{2},~R_{0}^{'}=r_{\Sigma}a_{0}^{'}$$

From the above solution for $\varepsilon > -1$, (see fig.2 and
fig.3) as before $t=t_{0}$ is the time of formation of
singularity. For $\varepsilon
> -\frac{2}{3}$, the singularity is covered by the event horizon
i.e., black hole will form, while for $-1 < \varepsilon <
-\frac{2}{3}$, the collapse will form a naked singularity with
zero mass at $t=t_{0}$ i.e., the mass of a black hole decreases
due to dark energy accretion and is
consistent with the claim of Babichev et al [16]. \\

As for equation of state $p=-\rho$ (i.e.,  $\varepsilon =-1$) the
solution corresponds to de sitter space so it is not of much
interest in the present context.\\

For $\varepsilon < -1$ i.e., when the matter in the brane is
purely in the form of dark energy the behavior of  both the
physical and the geometrical quantities of interest at the time
limits $T=T_{0}+\sqrt{\frac{A_{1}}{a^{3}}}~(=T _{s})$ and
$T\rightarrow
\infty$ are as follows:\\
$$T  \rightarrow T _{s}: ~~ a \rightarrow \infty,~~  \rho
\rightarrow \infty, ~~  \dot{R}(T) \rightarrow -\infty,~~ M(T)
\rightarrow \infty$$ $$T \rightarrow \infty:~~  a \rightarrow 0,~~
\rho \rightarrow 0,~~ \dot{R}(T) \rightarrow 0,~~ M(T) \rightarrow
0.$$

Thus if the cloud is untrapped initially, it will remain so
throughout the process and neither black hole nor naked
singularity will form. However, if initially $\dot{R}^{2}>1$
(trapped) then the cloud will start untrapped at the instant given
by the real root of the equation
\begin{equation}
\left[\frac{2}{3} R_{0}^{'}(T-T_{0})\right]^{6(|\varepsilon|
-1)}=\left[a_{3}(T-T_{0})^{2}-A_{1}\right]^{6|\varepsilon| -4}
\end{equation}
Further, it is to be noted that with the collapsing process both
the total mass and energy density of the cloud decrease and
finally become zero in the limit $T\rightarrow \infty$, though it
$(T=\infty)$ is not a singularity of space time.\\

\section{\normalsize\bf{effect of a combination of dark matter and dark energy in the gravitational collapse}}

The function $Q(t)$ in the conservation equation $(7)$ and $(8)$
stands for the interaction between dark matter and energy.\\

{\bf STEP-I : when $Q=0$ i.e., no interaction:}\\

The integral of equations $(7)$ and $(8)$ give
\begin{equation}\left.
\begin{array}{llll}
\rho_{DM}=\frac{\rho_{0}}{a^{3}}
\\\\
\rho=\frac{\rho_{1}}{a^{3(1+\varepsilon)}}
\end{array}\right\}
\end{equation}

where $\rho_{0}$ and $\rho_{1}$ are positive integration constant.
If $\varepsilon<-1$ then both $a=0$ and $a=\infty$ correspond to
space-time singularity while for $\varepsilon \geq -1$, $a=0$ is
the only  space-time singularity. These expressions for
$\rho_{DM}$ and $\rho$ result an integral form of the field
equation $(4)$ as

\begin{equation}
\int
\frac{dy}{\sqrt{(\rho_{0}+\frac{\rho_{1}}{y^{2\varepsilon}})\{1+\frac{1}{2\lambda
y^{2}}(\rho_{0}+\frac{\rho_{1}}{y^{2\varepsilon}})\}}}=\Lambda
-\sqrt{\frac{3\kappa}{4}}~t
\end{equation}
where $y=a^{\frac{3}{2}}$ and $\Lambda$ is an integration
constant. This integration can not be evaluated in general. For
$\varepsilon =-1$ the scale factor gives

\begin{equation}
a^{3}=\frac{1}{\rho_{1}}\left[(\frac{2\lambda \rho_{0}}{\rho_{1}+2
\lambda}) \cosh ^{2}[\sqrt{\frac{\rho_{1}(\rho_{1}+2
\lambda)}{2\lambda}}\{\sqrt{\frac{3\kappa}{4}}~(t_{0}-t)+A\}]-\rho_{0}\right]
\end{equation}
with $A=\sqrt{\frac{2\lambda}{\rho_{1}(\rho_{1}+2 \lambda)}}
\cosh^{-1} \sqrt{\frac{(\rho_{1}+2 \lambda)}{2\lambda}}$~.\\

The expressions for other relevant parameters are
\begin{equation}
\dot{R}(T)=-\sqrt{\frac{\kappa}{3a}}\sqrt{\rho_{0}+\rho_{1}a^{3}}\sqrt{1+{\frac{1}{2\lambda
a^{3}}(\rho_{0}+\rho_{1}a^{3})}}
\end{equation}
\begin{equation}
M(T)=\frac{\kappa r_{\Sigma}}{6}(\rho_{1}+\rho _{0}
a^{3})[1+\frac{1}{2\lambda a^{3}}(\rho_{1}+\rho_{0}a^{3})]
\end{equation}
 We note that at $T=T_{0}$ the space-time will be singular and the limiting
value of different parameters are
\begin{equation}\left.
\begin{array}{llll}
t\rightarrow t_{0}: a\rightarrow 0, \rho_{DM}\rightarrow \infty,
\rho=\rho_{1}, \dot{R}(T) \rightarrow -\infty,
M(T)\rightarrow\infty
\\\\
t\rightarrow \infty:a\rightarrow \infty, \rho_{DM}\rightarrow 0,
\rho=\rho_{1}, \dot{R}(T) \rightarrow -\infty,
M(T)\rightarrow\infty
\end{array}\right\}
\end{equation}\\

As the energy density for dark energy is constant so it behaves as
cosmological constant. Also near the singularity $(a=0)$ the dark
matter dominates over the dark energy. In fact, from the equation
$(22)$ we note that for general $\varepsilon < 0$, the dark matter
approaches infinity faster than dark energy as the collapsing
process approaches to singularity  (see fig.4).\\\\

\begin{figure}
\includegraphics[height=2.7in]{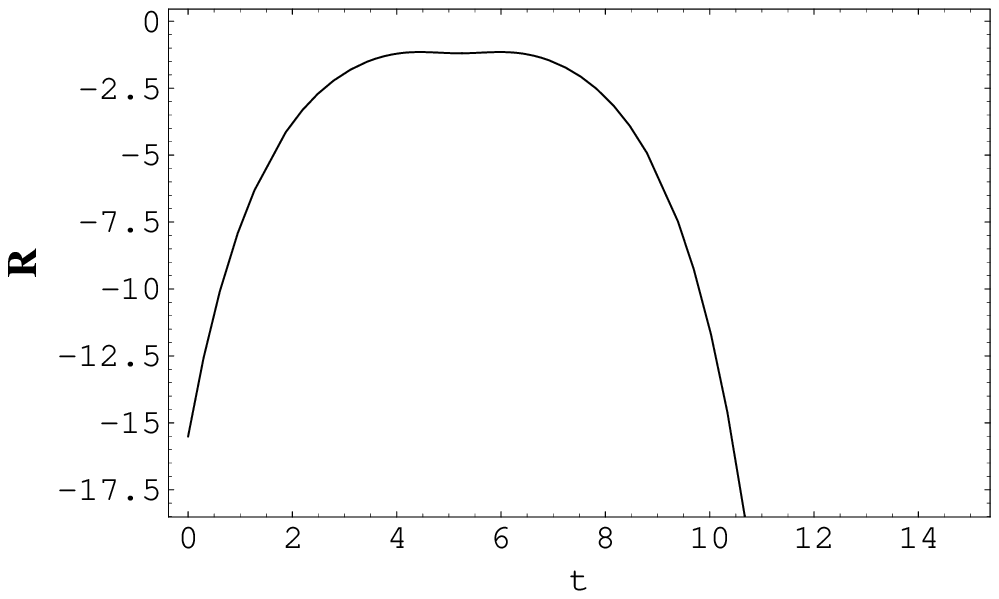}\\
\vspace{1mm} Fig.4\\

\vspace{5mm} Fig. 4 shows the variation of $\dot{R}$ with time in
eq. (25).  \hspace{2cm} \vspace{4mm}
\end{figure}

{\bf STEP II : $Q \neq 0$ i.e., Non Vanishing Interaction Between
Dark Matter And Dark Energy:}\\

Here according to Cai and Wang $[17]$ let us assume
\begin{equation}
\frac{\rho}{\rho_{M}}=B a^{3n}
\end{equation}
where $B (>0)$ and $n$ are arbitrary constants. Defining,
$\rho_{T}=\rho_{M}+\rho$, the total energy density, we have from
conservation equations (7) and (8)
\begin{equation}\left.
\begin{array}{llll}
\rho=\frac{B \rho _{2} a^{3n}}{a^{3} (1+B
a^{3n})(\frac{\varepsilon+n}{n})}
\\\\
\rho_{M}=\frac{\rho_{2}}{a^{3}(1+ B
a^{3n})^{(\frac{\varepsilon+n}{n})}}
\\\\
\rho_{t}=\frac{\rho_{2}}{[a^{3} (1+B
a^{3n})^{\frac{\varepsilon}{n}}]}
\end{array}\right\}
\end{equation}\\\
with $\rho_{2}$, an arbitrary positive constant. Then the field
equation $(4)$ can be written in the integral form as

\begin{equation}
\int\frac{(1+B z^{n})^{\frac{\varepsilon}{n}} dz}{\sqrt{z (1+B
z^{n})^{\frac{\varepsilon}{n}}+\frac{\rho_{2}}{2\lambda}}}=-\sqrt{3\kappa
\rho_{2}}~t +\Lambda_{1}
\end{equation}
where $z=a^{3}$ and $\Lambda_{1}$ is an integration constant. As
before the integral can only be evaluated for $\varepsilon =-1$
and $n=1$ and the solution is
\begin{equation}\left.
\begin{array}{llll}
a^{3}=\frac{1}{B}[D^{2} \cosh^{2}[\mu (t_{0}-t)+F]-1]
\\\\
\rho_{M}=\frac{\rho_{2}}{a^{3}}
\\\\
\rho=B \rho_{2 }
\\\\
\dot{R}(T)=-\frac{R_{2}\sqrt{1+Ba^{3}}}{a^{2}}[\frac{\rho_{2}}{2\lambda}+\frac{a^{3}}{D^{2}}]^{\frac{1}{2}}
\\\\
M(T)=\frac{M_{0}}{a^{3}}(1+Ba^{3})(\frac{\rho_{2}}{2\lambda}+\frac{a^{3}}{D^{2}})
\end{array}\right\}
\end{equation}\\\
where $$D=\sqrt{\frac{2\lambda}{2\lambda +B \rho_{2}}} ,F=\cosh
^{-1}(\frac{1}{D}) , R_{2}=r_{\Sigma}\sqrt{\frac{\kappa
\rho_{0}}{3}}, M_{0}=\frac{1}{2}r_{\Sigma}R_{2}^{2},
\mu=\sqrt{\frac{3\kappa
\rho_{2}B(2\lambda+B\rho_{2})}{8\lambda}}$$ From equation $(29)$,
the limiting form of the energy densities are
$$t \rightarrow t_{0}: a\rightarrow 0, \rho \sim a^{3(n-1)}, \rho_{M}\sim a^{-3}, \rho_{T}\sim a^{-3}
$$$$t \rightarrow  -\infty: a \rightarrow \infty, \rho \sim a^{-3(1+\varepsilon)}, \rho_{M}\sim a^{-3(1+\varepsilon +n)},
\rho_{T}\sim a^{-3(1+\varepsilon)}$$ Thus we see that $a=0$ is
always a singularity of space-time while $a=\infty$ is also
singular for $\varepsilon > -1$. Also for $n \geq 1$ near the
singularity $(a=0)$ the dark energy becomes insignificant and the
total matter is governed by dark matter. For $0<n<1$, though
$\rho$ and $\rho_{M}$ both approaches to infinity near the
singularity, but still the dark matter dominates over the
dark energy. \\

Further, using this solution the expression for the interaction is
given by (from equation $(7)$)
\begin{equation}
Q(t)=-3B (\varepsilon +n)(\frac{\dot{a}}{a})\frac {a^{3n}
\rho_{t}}{(1+B a^{3n})^{2}}
\end{equation}
which approaches to $-\infty$ at the singularity.\\

\section{\normalsize\bf{discussion and concluding remarks }}

The gravitational collapse of a homogeneous spherical cloud
consists of dark matter and dark energy has been considered here
in the background of brane world. To study the role of dark matter
and dark energy in the final outcome of gravitational collapse,
both of them have been considered separately as well as in
combination with or without interaction.\\

When the collapsing fluid is in the form of dark energy with
equation of state $p=\varepsilon \rho$, then the collapse will
lead to a black hole for $\varepsilon > - \frac{2}{3}$ while the
singularity will be naked for $-1 <\varepsilon< -\frac{2}{3}$.
However for $\varepsilon =-\frac{2}{3}$ the end state of collapse
will depend on the magnitude of
$\frac{r_{\Sigma}\rho_{1}}{2}\sqrt{\frac{\kappa}{6\lambda}}$. (If
this quantity is greater than unity then there will be black hole
otherwise there will be a naked singularity). Then the time of
formation of apparent horizon is given by $$t_{ah}=t_{0}
+\frac{1}{a_{1}}\left(A_{1}-\frac{1}{2R_{1}}\right)$$ This result
is distinct from Einstein gravity where the critical role is
played by $\varepsilon=-\frac{1}{3}$ and there is no restriction
on the parameters involved. \\

Then the section IV a combination of dark matter and dark energy
is considered as collapsing fluid. Due to the complicated form of
the equations involved solutions are obtained only for particular
value of the parameters. In step I solutions are obtained for
$Q=0$ (without interaction) for $\varepsilon =-1$ and solutions
for $Q \neq 0$ (with interaction) are presented in step II for
$\varepsilon =-1$ and $n=1$. In both cases, the solutions are
similar. At the time of singularity the collapsing cloud is
clearly trapped. But for regular initial data, the initial instant
of collapse should be chosen appropriately. As it is noted that
$\dot{R}\rightarrow - \infty$ both at $t\rightarrow -\infty$ and
$t=t_{0}$, so $\dot{R}$ has a finite maximum value (at the instant
$t=t_{max}(< t_{0})$). If this maximum value of $\dot{R}$ (which
depends on the brane tension $\lambda$) is chosen to be less than
unity (in magnitude) then any instant around $t_{max}$
may be taken as the starting epoch of the collapsing process.\\

Thus we conclude that the brane scenario has an effect on the
collapsing process. Finally, if the brane tension approaches to
infinity then one recovers Einstein gravity and the above results
reduce to those of Cai and Wang.\\\\

{\bf Acknowledgement:}\\\\
One of the authors (SC) is thankful to CSIR, Govt. of India for
providing a research project No. 25(0141)/05/EMR-II.\\

{\bf References:}\\
\\
$[1]$  L. Randall and R. sundrum, {\it Phys. Rev. Lett.} {\bf 83} (1999) 4690.\\
$[2]$  R.maartens, {\it Phys. Rev. D}  {\bf 62} (2000) 084023; T.
Shiromizu, K. Maeda,and M. sasaki,{\it Phys. Rev. D} {\bf 62}
(2000) 024012.\\
$[3]$ A. G. Riess etal. [Supernova Search Team collaboration] {\it
Astron. J.} {\bf 116} (1998) 1009; {\it Astrophys. J.} {\bf 607}
(2004) 665; S. Perlmutter etel [Supernova cosmology project collaboration] {\it Astrophys. J. Suppl.} {\bf 517}
(1999) 565.\\
$[4]$ C. L. Bennett etal., {\it Astrophys. J. Suppl.} {\bf 148}
(2003) 1; D. N. Spergel etal [WMAP Collaboration] {\it Astrophys.
J. Suppl.} {\bf 148} (2003) 175.\\
$[5]$  L. Verde etal, {\it Mon. Not. Roy Astron. Soc.} {\bf 335}
(2002) 432; E. Hawkins etal. , {\it Mon. Not. Roy Astron. Soc.}
{\bf 346} (2003) 78; K. Abazajian etal. [SDSS  collaboration ] {\it Phys. Rev. D}  {\bf 69} 103501 (2004).\\
$[6]$ V. Sahni and A. A. Starobinsky, {\it Int. J. Mod. Phys. D} {\bf 9} (2000) 373.\\
$[7]$  S. M. Carrpll, {\it Living Rev. Rel.} {\bf 4} (2001) 1.\\
$[8]$ P. J. E. Peebles and B. Ratra, {\it Rev. Mod. Phys.} {\bf 75} (2002) 559.\\
$[9]$  T. Padmanabhan, {\it Phys. Rep.} {\bf 380} (2003) 235.\\
$[10]$  V. sahni,``Dark Matter and Dark Energy"  {\it astro-ph}/ 0403324 (2004).\\
$[11]$  V. Sahni, ``Cosmological Surprises from Brane world models of Dark energy"  {\it astro-ph}/ 0502032 (2005).\\
$[12]$ Rong-Gen Cai and A. Wang, ``BlockHole Formation from
Collapsing Dark Matter in the Background of Dark Energy"
 {\it astro-ph}/0505136 (2005).\\
$[13]$ K. Maeda and D. Wands, {\it Phys. Rev. D}  {\bf 62} (2000) 124009.\\
$[14]$ M. Saski, T. Shiromizu and  K. Maeda {\it Phys. Rev. D}  {\bf 62} (2000) 024008.\\
$[15]$P. S. Joshi, {\it Global Aspects in Gravitation and Cosmology} (Clarendon, oxford, 1993).\\
$[16]$ E. Babichev, V. Dokuchave and Yu. Eroshenko, {\it Phys. Rev. Lett.}  {\bf 93} (2004) 021102 .\\
$[17]$ R, -G. Cai and A. Wang., {\it JCAP} 0503 (2005) 002.\\

\end{document}